\documentclass[aps,showpacs,preprint]{revtex4}
\usepackage{epsfig,amssymb,bm,graphics}
\newcommand*{\fs}[1]{#1\!\!\!/}
\begin{document}

\title{Dimuon production by laser-wakefield accelerated electrons}

\author{A.I.~Titov$^{a,b,c}$, B.~K\"ampfer$^{a,d}$ and H. Takabe$^c$}
 \affiliation{
 $^a$Forschungzentrum Dresden-Rossendorf, 01314 Dresden, Germany\\
 $^b$Bogoliubov Laboratory of Theoretical Physics, JINR,
  Dubna 141980, Russia\\
 $^c$Institute of Laser Engineering, Yamada-oka, Suita,
  Osaka 565-0871, Japan\\
 $^d$ Institut f\"ur Theoretische Physik, TU~Dresden, 01062 Dresden,
 Germany
 }

\begin{abstract}
  We analyze $\mu^+\mu^-$ pair production generated by
  high-energy electrons emerging from a laser-wakefield
  accelerator. The $\mu^+\mu^-$ pairs are created in a solid
  thick high-$Z$ target, following the electron accelerating
  plasma region.
  Numerical estimates are presented for electron beams
  obtained presently in the LBL TW laser experiment~\cite{C2}
  and possible future developments.
  Reactions induced by the secondary
  bremsstrahlung photons dominate the dimuon production.
  According to our estimates, a 20 pC electron bunch with energy of 1 (10) GeV
  may create about 200 (6000) muon pairs.
  The produced $\mu^\pm$ can be used in studying various
  aspects of  muon-related physics in table top installations.
  This may be considered
  as an important step towards the investigation of more complicated
  elementary processes induced by laser driven electrons.
\end{abstract}


\maketitle

\section{introduction}

The possibility to produce strong electric fields
of the order of $10-100$~GV/m
with present laser facilities
is a great advantage
for laser-wakefield accelerators~\cite{Tajima1979}
which allows, in principle, to construct  compact
accelerating devices for particle and nuclear physics.
The successful production
of high-quality electron beams in such laser-driven
accelerators with electron energies of the order of
1 GeV has been reported recently~\cite{C2,C3,C4}.
Electron beams with energies exceeding $1$~GeV are interesting
for many applications in particle physics, such as investigating the properties
and production mechanisms of
vector and exotic scalar mesons in
photo/electroproduction~\cite{Vector,Scalars},
excitation of baryon resonances with the aim
of studying their properties and search for
missing resonances~\cite{Oh2001}, strangeness
photoproduction~\cite{Strangeness} etc.
Another subject is related to the neutrino physics.
For example, neutrino oscillations need
two types of neutrinos, at least. They may be obtained in muon decays
$\mu^+\to e^+ +\nu_e +\bar \nu_\mu$ and
$\mu^-\to e^- + \bar \nu_e +\nu_\mu$, where muons and electrons
neutrinos (or antineutrinos) are produced in equal parts.
Therefore, it is interesting to estimate
whether the high-energy laser-driven electrons
can produce a sizeable amount of muon pairs for future applications.

Recall that the present generation of the high-energy neutrino beams
are made by decays of charged pions and kaons
in flight in a long decay channel.
In this case, the neutrino (antineutrino) beams mostly
consists of $\nu_\mu\, (\bar \nu_\mu)$ with rather small admixture
of electron neutrinos.
The idea to use a storage ring of
muons as a source of high-energy and high-intensity neutrino beam
has been discussed in Refs.~\cite{Geer1998,Geer2002}.
It is assumed that the muons are produced in two-body decays
$\pi^+\to \mu^++\nu_\mu$ and $K^+\to \mu^++\nu_\mu$ and then
stored in a ring with subsequent decay into an electron
and two neutrinos. The pions (kaons) in turn, are produced in proton-proton
and (or) in proton-nucleus collisions with high-intensity
proton beams~\cite{ISS}. Together with neutrino oscillation,
such high-intensity muon sources
may be used in studying other fundamental problems of lepton physics,
say the search for lepton flavor violation~\cite{LFV} and
the measurement of the muon's anomalous magnetic moment~\cite{Mu_g2}.
The idea to use a laser-driven proton beam
for investigating different aspects of the
neutrino oscillations was discussed
for the first time in Ref.~\cite{Bulanov}.

The aim of present paper is to analyze the possibility
of muon pair creation in the interaction of high-energy
laser driven electrons within a heavy (high-$Z$) target
in a table top configuration.
The electromagnetic sources of the $\mu^+\mu^-$
(dimuon) production are described by the following elementary processes
\begin{eqnarray}
 \gamma +A&\to& A+ \mu^+\mu^-\, ,\label{E1a}\\
 e+A&\to& e'+A+\mu^+\mu^-\, .\label{E1b}
\end{eqnarray}
In the first case (1) the dimuons  are produced
in the interaction of real (bremsstrahlung) photons within the
electric field of the high-$Z$ target nuclei. This is an analog of
well known Bethe-Heitler process of the electron-positron production.
In the second case (2), the dimuons are produced in the interaction of
high-energy electrons with nuclei (so called trident process).
These two reactions are depicted in
Fig.~1.
In some sense
   \begin{figure}[h!]
    \includegraphics[width=0.4\columnwidth]{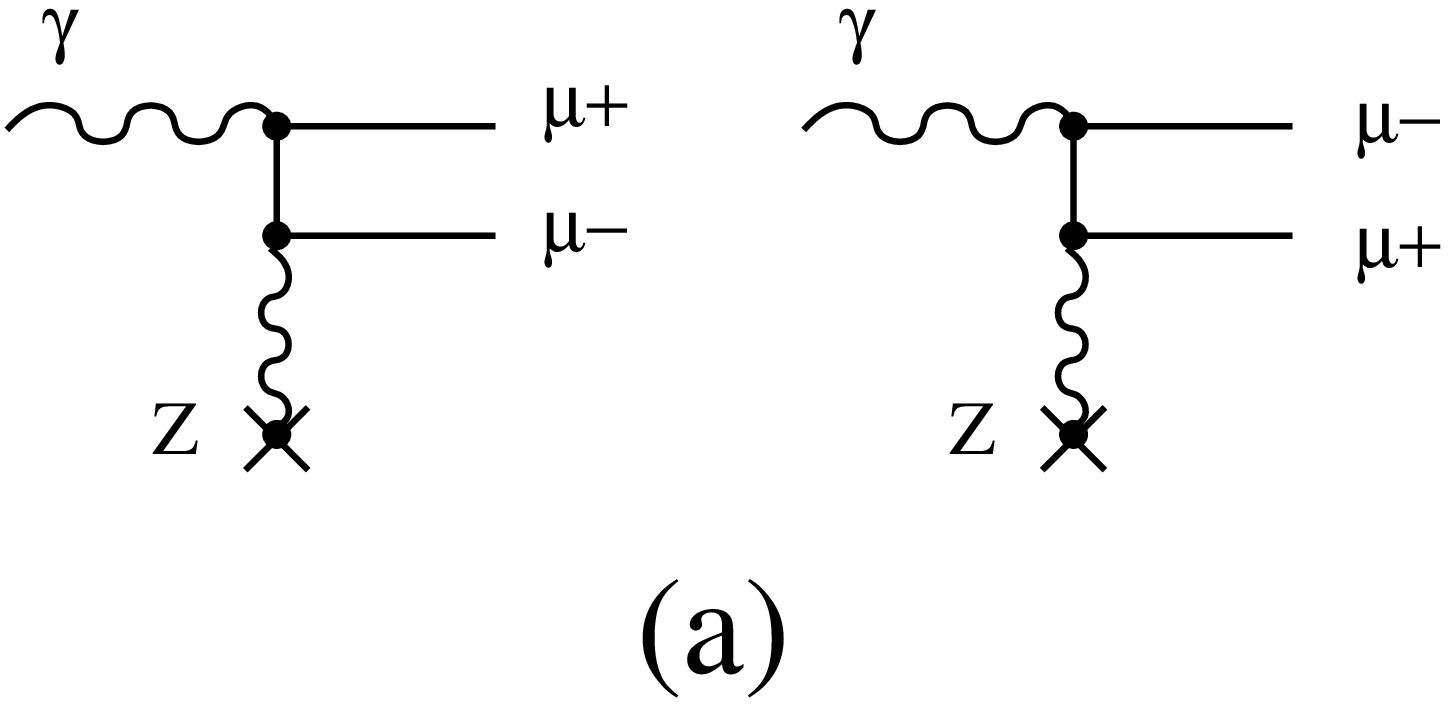}\qquad\qquad
        \includegraphics[width=0.4\columnwidth]{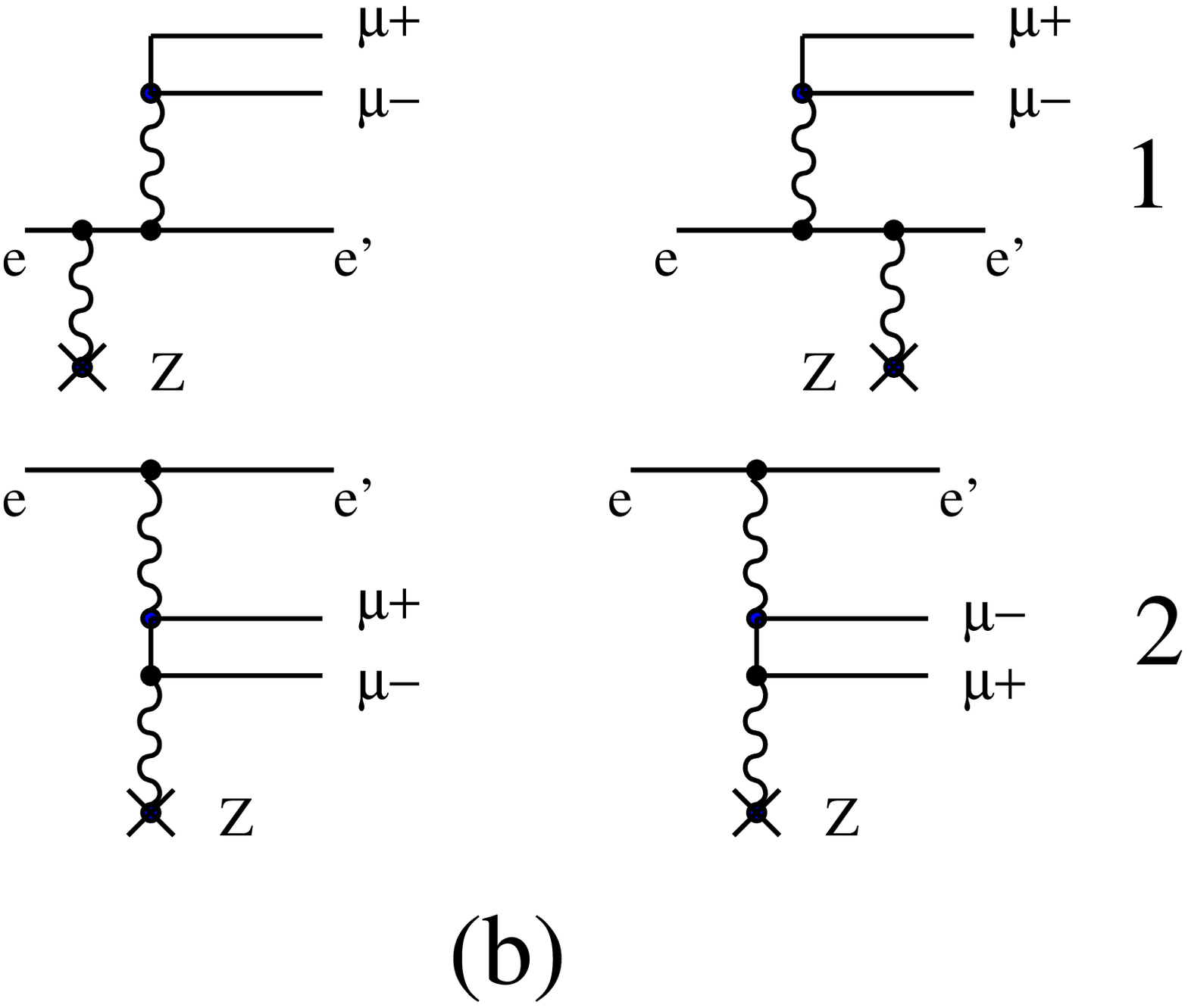}
   \caption{\small{
   Diagrammatic representation of
   dimuon production processes in electromagnetic interactions.
   (a) Bethe-Heitler process $\gamma +A\to A+ \mu^+\mu^-$.
   (b) Electron induced reaction $e+A\to e+A+\mu^+\mu^-$.
\label{Fig:1} }}
  \end{figure}
these reactions are similar to the dielectron production by
laser-driven relativistic electrons at few dozen MeV analyzed
by Nakashima and Takabe~\cite{NT2002} and other authors
(see, for example,
Ref.~\cite{Karsch1999}). Therefore, our investigation
may be considered as a continuation of these studies,
but with the focus on dimuon production.
For other avenues towards laser-generated muon pair production
cf. Refs.~\cite{Ruf2009,Muller2008}.
Dimuon production in electron-positron-photon plasmas was
estimated in~\cite{Raf}.

Our paper is organized as follows. In Sect.~II,
we recall the relevant elementary electromagnetic processes.
Estimates of the dimuon yield
are presented in Sect.~III, where the energy loss and
attenuation of the primary electron beam and the secondary
photon bremsstrahlung spectrum are accounted for.  Hadronic processes
are briefly discussed in Sect.~IV. Our conclusions
can be found in Sect.~V.

\section{elementary cross sections}

The differential cross section of the dimuon production in
the reaction Eq.~(\ref{E1a}) reads
\begin{eqnarray}
 d\sigma^{\gamma\,A}=\frac{Z^2\alpha^3}{4\pi}
 \,\frac{|{\bf q}|}{|{\bf k}|}\,|T^{\gamma\,A}|^2\,
 \sqrt{M^2_{\mu^+\mu^-}-4\mu^2}\,dM_{\mu^+\mu^-}\,d\cos\theta_q
 d\Omega_{\mu^+},
 \label{E2}
 \end{eqnarray}
where ${\bf k}$ and ${\bf q}$ are the spatial  parts of the
incoming photon momentum and the total momentum of the outgoing
muon pair $q=p_{\mu^+}+p_{\mu^-}$ with
the invariant mass of the muon pair
$M_{\mu^+\mu^-}^2=q^2$; $\mu$ denotes the muon mass,
 $\theta_q$ is the polar angle of the direction of flight of dimuon,
 $\Omega_{\mu^+}$ is the solid angle of the
direction of the velocity of the $\mu^+$ meson in the rest frame of the dimuon;
$Z$ and $\alpha$ stand for the nuclear charge and the fine structure
constant ($\alpha=1/137$), respectively; $T^{\gamma\,A}$
is the invariant amplitude and averaging over photon polarization and
summation over $\mu^\pm$ spin states is provided in $|T^{\gamma\,A}|^2$.
The quantization axis $z$ is chosen along velocity
of incoming photon, and the production plane is defined by the vectors
${\bf k}$ and ${\bf q}$. The differential cross section depends
on the initial energy and four kinematical variables,
which define the final state.

In case of the $e+A\to e'+A+\mu^+\mu^-$ reaction Eq.~(\ref{E1b})
we have relevant two sub-processes.
One corresponds to the electron scattering off the nucleus.
In the second case, one of the outgoing muons interacts
with the atomic nucleus. These two cases are marked in Fig.~1~(b)
as "1" and "2", respectively.
The total amplitude is the coherent
sum of all four amplitudes.   The differential cross section
for this reaction depends on seven kinematical variables,
which define the final state of the
outgoing leptons and recoil nucleus,
and it has the form
 \begin{eqnarray}
 d\sigma^{e\,A}=\frac{Z^2\alpha^4}{16\pi^3}
 \,\frac{|{\bf p}_{e'}|}{|{\bf p}_e|}\,|T^{e\,A}|^2\,
 \sqrt{M^2_{\mu^+\mu^-}-4\mu^2}\,|{\bf q}|dE_q\,dM_{\mu^+\mu^-}
 \,{d\cos\theta_{e'}}\,{d\Omega_{q}}\,{d\Omega_{\mu^+}},
 \label{E3}
 \end{eqnarray}
where ${\bf p}_e$ and ${\bf p}_{e'}$ are three  momenta of the incoming and
outgoing electrons, respectively, $\theta_{e'}$ is the polar angle
of the direction of flight of the outgoing electron,
$E_q=\sqrt{{\bf q}^2+M_{\mu^+\mu^-}^2}$ is the
energy of the muon pair;
 $T^{e\,A}$ is the invariant amplitude and averaging over initial electron spin
 projections and summation over the final fermion spin states
 is understood in  $|T^{e\,A}|^2$;
the quantization axis ($\bf z$) is along velocity of the incoming electron,
and ${\bf y}=[{\bf p_e}\times {\bf p_{e'}}] /|{\bf
p_e}||{\bf p_{e'}}|$.

The invariant amplitudes for reactions in Eqs.~(\ref{E1a}) and (\ref{E1b})
are calculated in the lowest order of perturbation theory analog to the
electron-positron production in $\gamma\,A$ and $e\,A$ interactions~\cite{AB}.
The amplitude for the $\gamma\,A$ reaction reads
\begin{eqnarray}
T^{\gamma\,A}=-\frac{1}{\kappa_\gamma^2}
\bar u_{\mu^-}\left[
 \gamma_0\frac{{\fs g}_1+\mu}{g_1^2-\mu^2}\,\gamma_\alpha
-\gamma_\alpha\frac{{\fs g}_2-\mu}{g_2^2-\mu^2}\,\gamma_0
\right]\,v_{\mu^+}\,\varepsilon_\gamma^\alpha,
\label{A1}
\end{eqnarray}
where $u$ and $v$ are the Dirac spinors of the outgoing $\mu^-$ and $\mu^+$ muons, respectively, $\varepsilon_\gamma$ is the polarization vector of incoming photons,
$g_1=k-p_{\mu^+}$, $g_2=k-p_{\mu^-}$, and
$\kappa_\gamma=k-q$. The notation $\fs p$ means
the four product $\gamma_\alpha p^\alpha$, where $\gamma_\alpha$ ($\alpha=0,1,2,3$)
are Dirac's $\gamma$ matrices.

Correspondingly, for the $e\,A$ reaction we have
\begin{eqnarray}
T^{e\,A}=T^{e\,A}_1+T^{e\,A}_2
\nonumber
\end{eqnarray}
with
\begin{eqnarray}
T^{e\,A}_1&=&-\frac{1}{\kappa_e^2\,q^2}
\left(\bar u_{e'}
\left[
 \gamma_0\frac{{\fs p}_1+ m_e}{p_1^2-m_e^2}\,\gamma_\alpha
-\gamma_\alpha\frac{{\fs p}_2+ m_e}{p_2^2-m_e^2}\,\gamma_0
\right]\,u_{e}\right)
\, \left(\bar u_{\mu^-}\gamma^\alpha\,v_{\mu^+}\right)~,
\label{A2}\\
T^{e\,A}_2&=&-\frac{1}{\kappa_e^2\,{q'}^2}
\left(\bar u_{e'}\,\gamma^\alpha\,u_{e}\right)\,
\left(\bar u_{\mu^-}\left[
 \gamma_0\frac{{\fs p}_3+\mu}{p_3^2-\mu^2}\,\gamma_\alpha
-\gamma_\alpha\frac{{\fs p}_4-\mu}{p_4^2-\mu^2}\,\gamma_0
\right]\,v_{\mu^+}\right),
\label{A3}
\end{eqnarray}
where $q'=p_e-p_{e'}$, $p_1=p_e-q$, $p_2=p_{e'}+q$,
$p_3=q'-p_{\mu^+}$, $p_4=q'-p_{\mu^-}$, and $\kappa_e=q'-q$.

The effects of nuclear size and higher-order
Coulomb corrections are small~\cite{Roche1971};
the effects of the atomic electron
screening~\cite{Schiff,BA1972} are neglected here
as the production processes happen in the high-field region
near the target nuclei.
The corresponding cross sections Eqs.~(\ref{E2}) and (\ref{E3})
are evaluated numerically without any additional approximation.

   \begin{figure}[h!]
       \includegraphics[width=0.3\columnwidth]{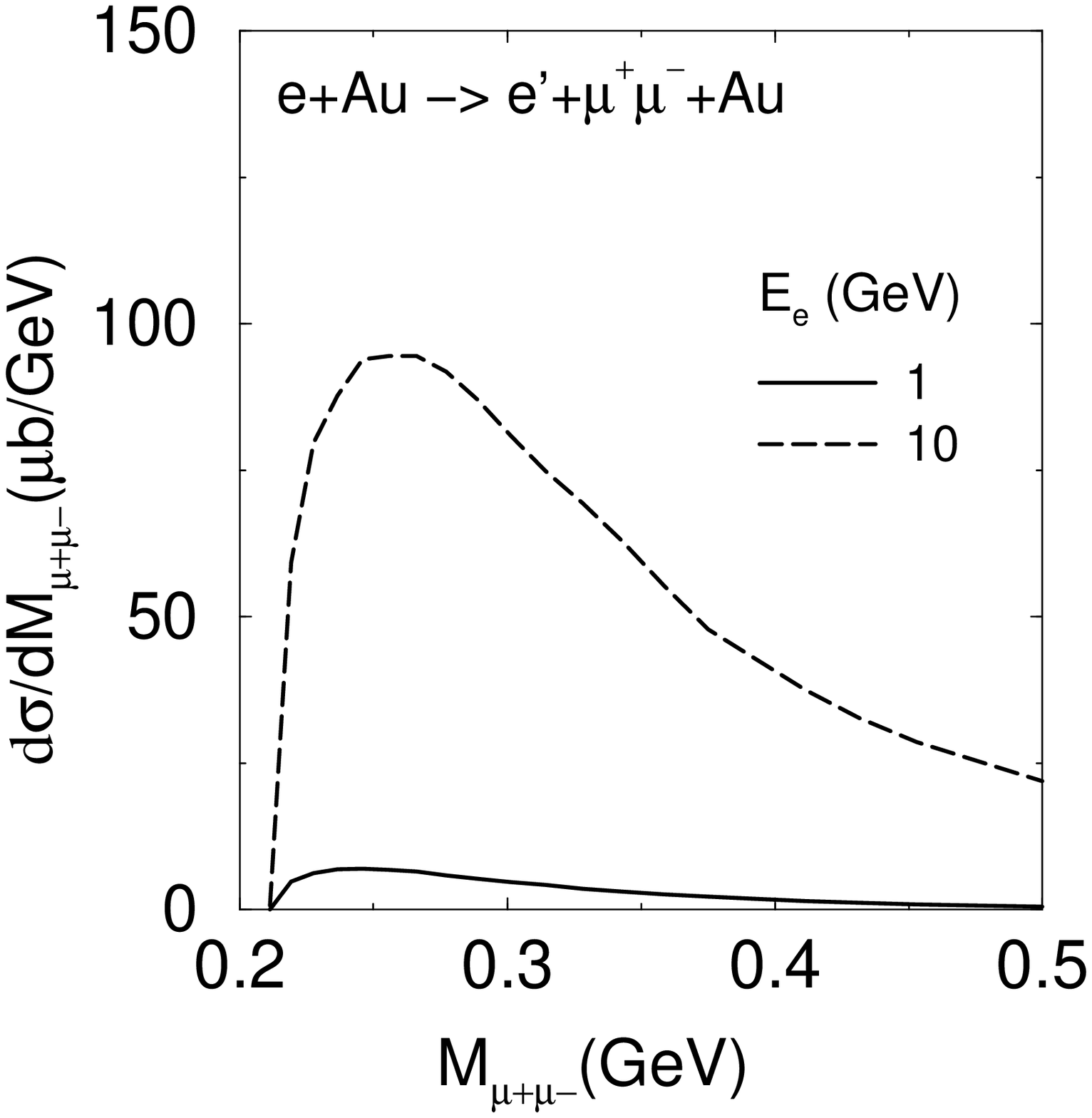}\qquad\qquad
        \includegraphics[width=0.3\columnwidth]{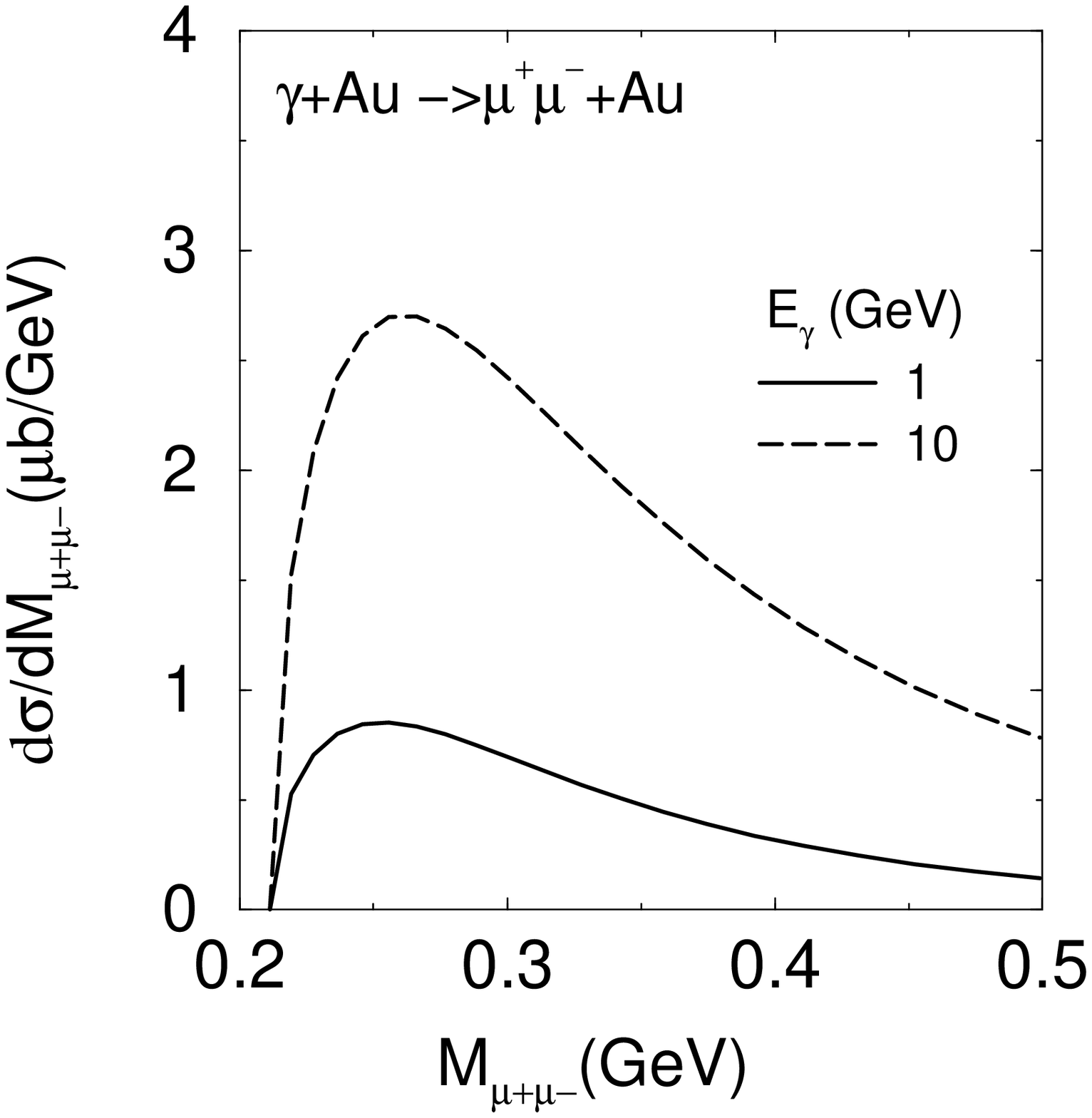}
   \caption{\small{
   Differential cross sections of dimuon production in
   $e\, A$ (left panel) and   $\gamma\, A$ (right panel)
   interactions as a function of the invariant dimuon mass
   $M_{\mu^+\mu^-}$. The solid and dashed curves correspond
   to initial energies of 1 and 10 GeV, respectively.
   \label{Fig:2} }}
  \end{figure}

The invariant-mass distributions for $\gamma A$ and $e A$  reactions are shown
in Fig.~{\ref{Fig:2}}. In this case, the corresponding
cross sections are calculated as a function of the invariant
mass $M_{\mu^+\mu^-}$ integrating over the other variables in
Eqs.~(\ref{E2}) and (\ref{E3}).
All calculations are done for a gold target.
The initial electron energies
are chosen to be 1 and 10 GeV.
For 1 GeV we follow the laser accelerator conditions of Ref.~\cite{C2}.
The case of 10 GeV may be considered as prediction for future
high-energy table top electron accelerators
being under consideration now~\cite{HEEA1,HEEA2,HEEA3}.
For simplicity, in case of the $\gamma\,A$ reaction, the elementary
cross sections are calculated at $E_\gamma=1$ and 10~GeV.
Below, for the estimate of the dimuon
yield, we fold the elementary cross section with the bremsstrahlung photon distribution
for all kinematically allowed photon energies.

   \begin{figure}[h!]
    \includegraphics[width=0.3\columnwidth]{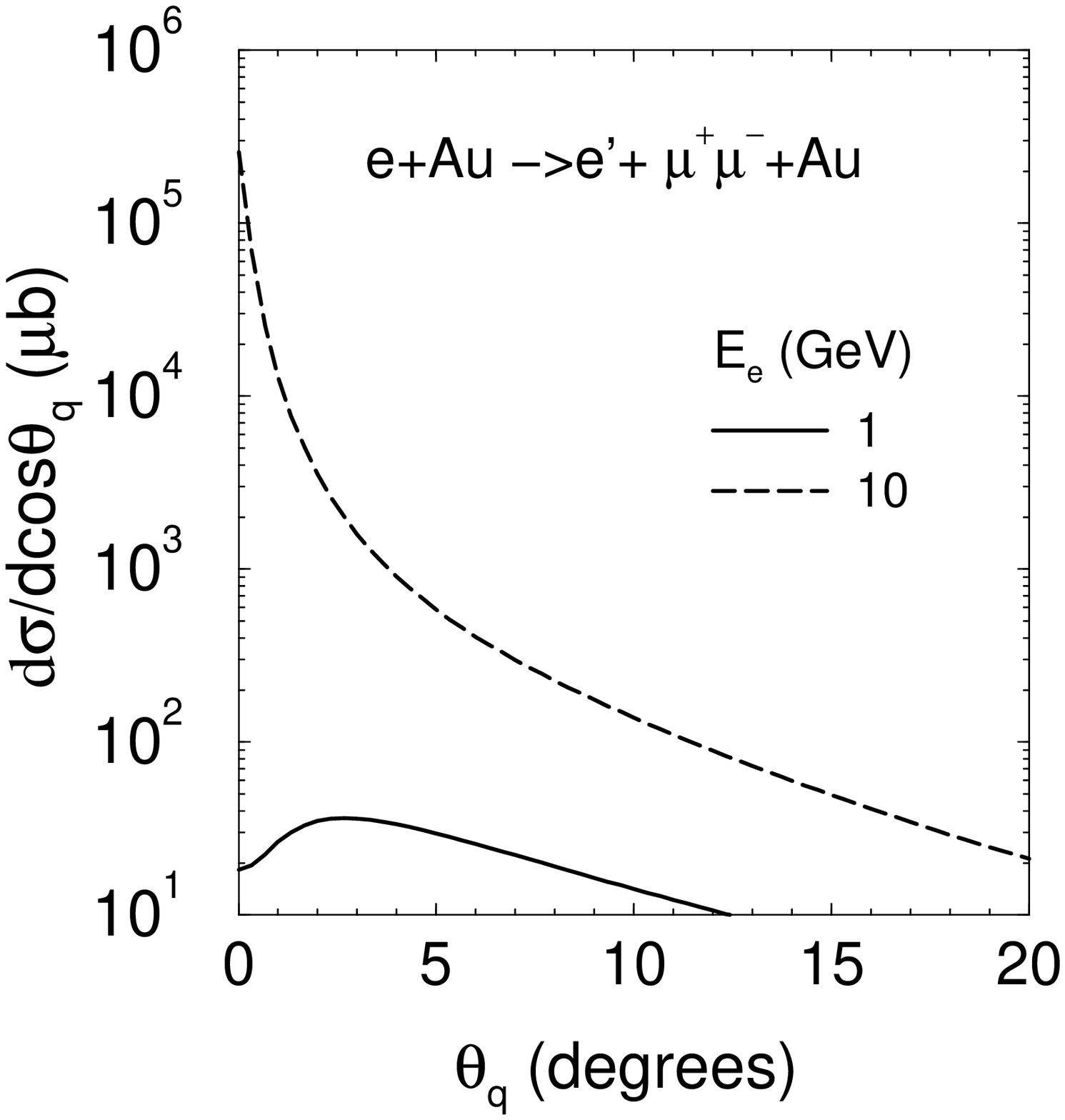}\qquad\qquad
        \includegraphics[width=0.3\columnwidth]{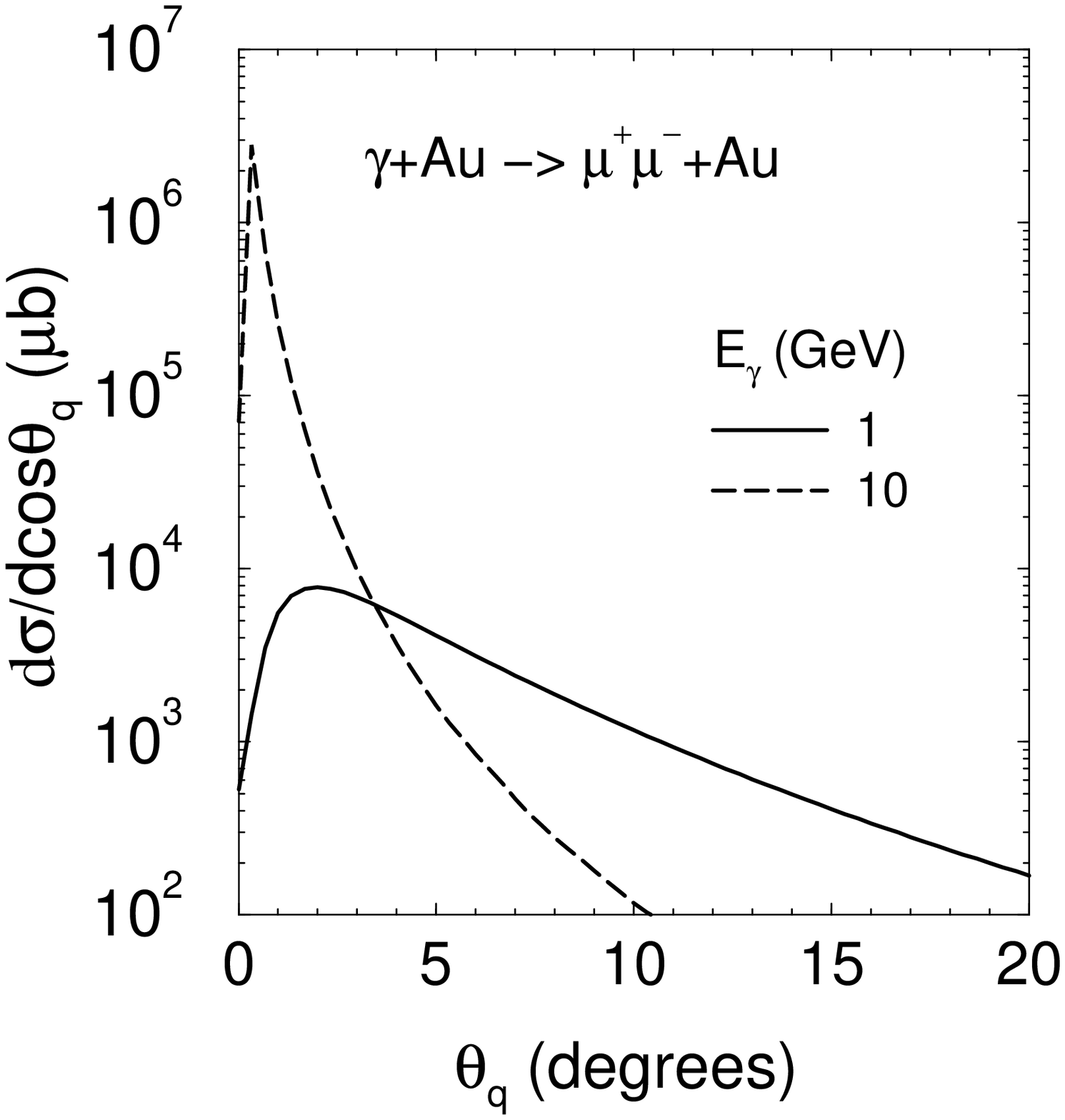}
   \caption{\small{
   Differential cross sections of dimuon production in
   $e\, A$ (left panel) and  $\gamma\, A$ (right panel)
   reactions as a function of polar angle $\theta_q$.
   The solid and dashed curves correspond
   to initial energies 1 and 10 GeV, respectively.
   \label{Fig:3} }}
   \end{figure}

Fig.~\ref{Fig:2} demonstrates  that the cross section
of the electron induced reaction is smaller by  two order
of magnitude than the corresponding cross section
in the photon induced reaction.
This is mainly because of an additional vertex with additional factor $\alpha$.
Both distributions exhibit a maximum slightly above $\sim2 \mu$.
This means that the relative kinetic energy
of the two muons in a pair is small,
i.e. they flight close to each other.

   \begin{figure}[h!]
    \includegraphics[width=0.3\columnwidth]{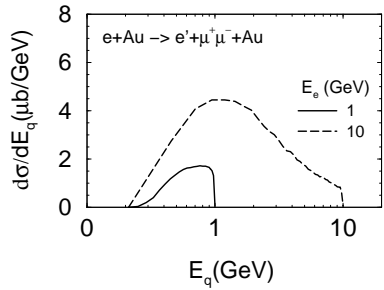}
   \caption{\small{
   The differential cross section for electron
   induced muon pair production
   reaction as a function of dimuon energy $E_q$.
   The solid and dashed curves correspond
   to initial energies 1 and 10 GeV, respectively.
   \label{Fig:4} }}
   \end{figure}

   The differential cross sections as a function of the polar
   angle of the direction of flight of the muon pair $\theta_q$
   integrated over the other variables are shown in Fig.~\ref{Fig:3}.
   One can see that the cross sections are peaked
   at the forward direction, especially for $E_{e,\gamma}=10$~GeV.
   The knowledge of the spatial structure
   of the outgoing muon flux is important for the design
   of the devices for muon sources.

   The differential cross section of the reaction $e+A\to e'+\mu^+\mu^-+A$
   as a function of the dimuon energy $E_q$ is presented
   in Fig.~\ref{Fig:4}. One can see a wide spread  of this distribution
   with a maximum around 1 GeV for initial electron energy of 10 GeV,
   while for 1 GeV electrons the dimuon energy displays a plateau above
   0.6 GeV till the kinematical limit. Since the kinetic
   energy of the relative motion of $\mu^+$ and $\mu^-$ within a pair is small
   $(E_{\mu^\pm}\simeq E_q/2)$,
   the energy distribution of the individual muons may be approximated as
   $d\sigma/dE_{\mu^+}\simeq d\sigma/dE_{\mu^-}\simeq d\sigma/dE_{q}\vert_{E_{\mu^\pm}=E_q/2}$.

   \begin{figure}[h!]
    \includegraphics[width=0.3\columnwidth]{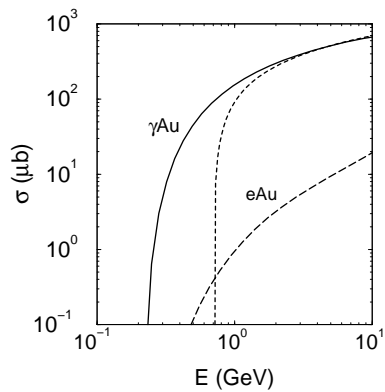}
   \caption{\small{
    The total cross section of the dimuon production in $\gamma\, A$ (solid curve)
     and $e\, A$ (dashed curve)
   reactions as a function of initial energy. The dotted curve corresponds
   to the high-energy approximation Eq.~(\protect\ref{URL}).
  \label{Fig:5} }}
  \end{figure}
 The total cross sections of the dimuon production
 as a function of the initial energy are shown in
 Fig.~\ref{Fig:5}. One can see a steep increase of the
 cross sections with energy.
 We note that in the ultra-relativistic case with $E_\gamma\gg\mu$ the total
 cross section of the reaction $\gamma A\to \mu^+\mu^-A$ may be described
 by~\cite{AB}
 \begin{eqnarray}
 \sigma\simeq\frac{28}{9}Z^2_A\alpha {r^\mu_0}^2
 \left( \ln \frac{2E_\gamma}{\mu}-\frac{109}{42}\right)~,
 \label{URL}
 \end{eqnarray}
 where $r^\mu_0=\alpha/\mu$ is the classical muon radius.
 The corresponding cross section is shown in
 Fig.~\ref{Fig:5} by the dotted curve
 showing that the ultra-relativistic behavior
 sets in at $E_\gamma\gtrsim 3$~GeV.

 The total cross sections  for
 $\gamma\, A$ and $e\, A$ reactions reach 300 $\mu$b and 3 $\mu$b,
  respectively,
 at $E_{\gamma(e)}\simeq 2$~GeV. Similarly to
 the differential cross sections, the difference
 between the two reactions is about two orders of magnitude,
 mainly due to an additional power of $\alpha$ in
 the $e\,A$ cross section.

\section{Dimuon yields}

Using the above elementary cross sections one can estimate the dimuon yield
for given electron beam and target properties.
For the former ones we use the conditions of
electron beams as reported for the laser-wakefield  accelerator in Ref.~\cite{C2}.
The electron energy is about 0.5 $-$ 1 GeV and the electron flux is
about $20$~pC which corresponds to $N^e_0\simeq1.248\times 10^8$ electrons
in a bunch. In our estimates we assume the same flux for electron energies
up to 10 GeV.  We consider a gold target with thickness of
$L=0.1-1$~cm.

Strictly speaking, the particle production in interactions of
the  high-energy electrons with heavy target nuclei must be evaluated
by transport-kinetic models
(see, for example, Refs.~\cite{NT2002,Karsch1999}).
However, for a first qualitative estimate of the dimuon yield one can use an
analytical approach, similar to that developed in Ref.~\cite{Sch1997}.

Consider first the dimuon yield which stems  from elementary
$eA\to e'+\mu^+\mu^-+A$ reactions.  It may be expressed as
\begin{eqnarray}
 d N^{\mu^+\mu^-}=
 \frac{N_A\,\rho_A}{A}\,\int\limits_0^L\,dl\,
  N^e(l)\,d\sigma^{eA\to e' \mu^+\mu^-A}(E_e(l)),
 \label{E4}
 \end{eqnarray}
where $A$ is the atomic weight, $N_A$ is Avogadro's number,
$\rho_A$ denotes the target density,
$d\sigma^{eA\to e'\mu^+\mu^- A}$ is the elementary cross section of the dimuon
production discussed above.
For the sake of simplicity, we neglect the energy
spread in the bunch taking the electron energy at the
central positions of the energy distribution.
This seems to be reasonable, because
the energy spreading reported in~\cite{C2}
is less than 100~MeV at $E_e\simeq1$~GeV.

Propagating through the target material the electron
beam loses its intensity and energy.
The energy loss is described by~\cite{AB}
\begin{eqnarray}
\frac{dE_e}{dl}=-E_e
\frac{N_A\,\rho_A\,\alpha\,Z_A^2{r^e_0}^2}{A}
\left(4\ln\frac{183}{Z_A^{1/3}}+\frac29
\right)
\label{E41}
\end{eqnarray}
with $r^e_0=\alpha/m_e\simeq 2.18\times10^{-13}$~cm.
For the gold target with $A=197$ and $Z_A=79$ this leads to the formula
\begin{eqnarray}
E_e(l)\simeq E_e^0\exp(-l/l_0)
\label{E42}
\end{eqnarray}
with $l_0\simeq0.513$~cm.
The electron beam absorption in the target may be accounted for by
a linear dependence
\begin{eqnarray}
N^e(l)\simeq N^e_0(1-l/L_{\rm max}(E_e^0))~,
\label{E43}
\end{eqnarray}
where $N^e_0$ is the initial beam intensity and
$L_{\rm max}$ stands for the maximum distance traveled by the high-energy electrons.
For the gold target one can use $L_{\rm max}\simeq 1.44$~cm at
$E_e=0.5$~GeV which may be extrapolated as
$L_{\rm max}(E)\simeq 0.23\,E^{0.3}$~cm with $E$ in GeV~\cite{Babichev}.
Both, the decrease of the beam intensity and the energy loss lead
to a decrease of the effectiveness of the dimuon production in the target
material. However, the effect of the energy loss is significantly greater
than the effect of intensity depletion since
 the scale parameter $l_0$ in Eq.~(\ref{E42})
is much smaller than $L_{\rm max}$.

\begin{figure}[h!]
   \includegraphics[width=0.3\columnwidth]{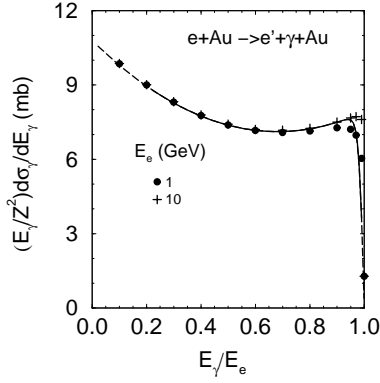}
   \caption{\small{
   The scaled bremsstrahlung cross section
   $(E_\gamma/Z^2)d\sigma_\gamma(E_e,E_\gamma)$
   as a function of $E_\gamma/E_e$ for two electron energies
   1~GeV (solid curve) and 10~GeV (dashed curve).
   The symbols represent the data compilation of Ref.~\protect\cite{BrData}.
\label{Fig:6} }}
  \end{figure}

The yield of dimuons, produced by secondary real photons created at the distance $l$
behind the target front side,
reads
\begin{eqnarray}
 d N^{\mu^+\mu^-}=\frac{N_A\,\rho_A}{A}\,
 \int\limits_0^L\,dl\int dE_\gamma \frac{dN_\gamma(l)}{dE_\gamma} \,(L-l)
 \,d\sigma^{\gamma A\to \mu^+\mu^-A}(E_\gamma)~,
 \label{E6}
 \end{eqnarray}
  where $d\sigma^{\gamma A\to \mu^+\mu^-A}$ is the above cross section of the dimuon
  production in the elementary $\gamma\,A$ interaction, and
  $dN_\gamma(l)/dE_\gamma$
  denotes the distribution of bremsstrahlung photons with energy $E_\gamma$ at distance $l$
 \begin{eqnarray}
 \frac{dN_\gamma(l)}{dE_\gamma}= \frac{N_A\,\rho_A}{A}
 N^e(l)\,\frac{d\sigma_\gamma(E_e(l),E_\gamma)}{dE_\gamma}~,
 \label{E7}
 \end{eqnarray}
where $N^e(l)$ is defined by Eq.~(\ref{E43}) and the current electron energy
$E_e(l)$ from Eq.~(\ref{E42}) is to be used.
For the angular-integrated bremsstrahlung cross section
$d\sigma_\gamma(E_e,E_\gamma)/dE_\gamma$
we adopt a parametrization motivated
by the analytical expression of Ref.~\cite{AB}
\begin{eqnarray}
\frac{E_\gamma}{Z_A^2}\frac{d\sigma_\gamma}{dE_\gamma}=
\frac{4\alpha {r^e_0}^2F(E_e,E_\gamma)}{E_e^2}
\left((E_e^2 +E_{e'}^2 -\frac23E_eE_{e'})\ln183Z_A^{-1/3}
+\frac{E_eE_{e'}}{9}
\right)~,
\label{E71}
\end{eqnarray}
where $E_{e'}=E_e-E_\gamma$ and $F(E_e,E_\gamma)=0.91\tanh(E_{e'}/0.02E_e)$.
In Fig.~\ref{Fig:6} we show the bremsstrahlung cross section for two
energies together with the compilation of Ref.~\cite{BrData}.

\begin{figure}[h!]
   \includegraphics[width=0.3\columnwidth]{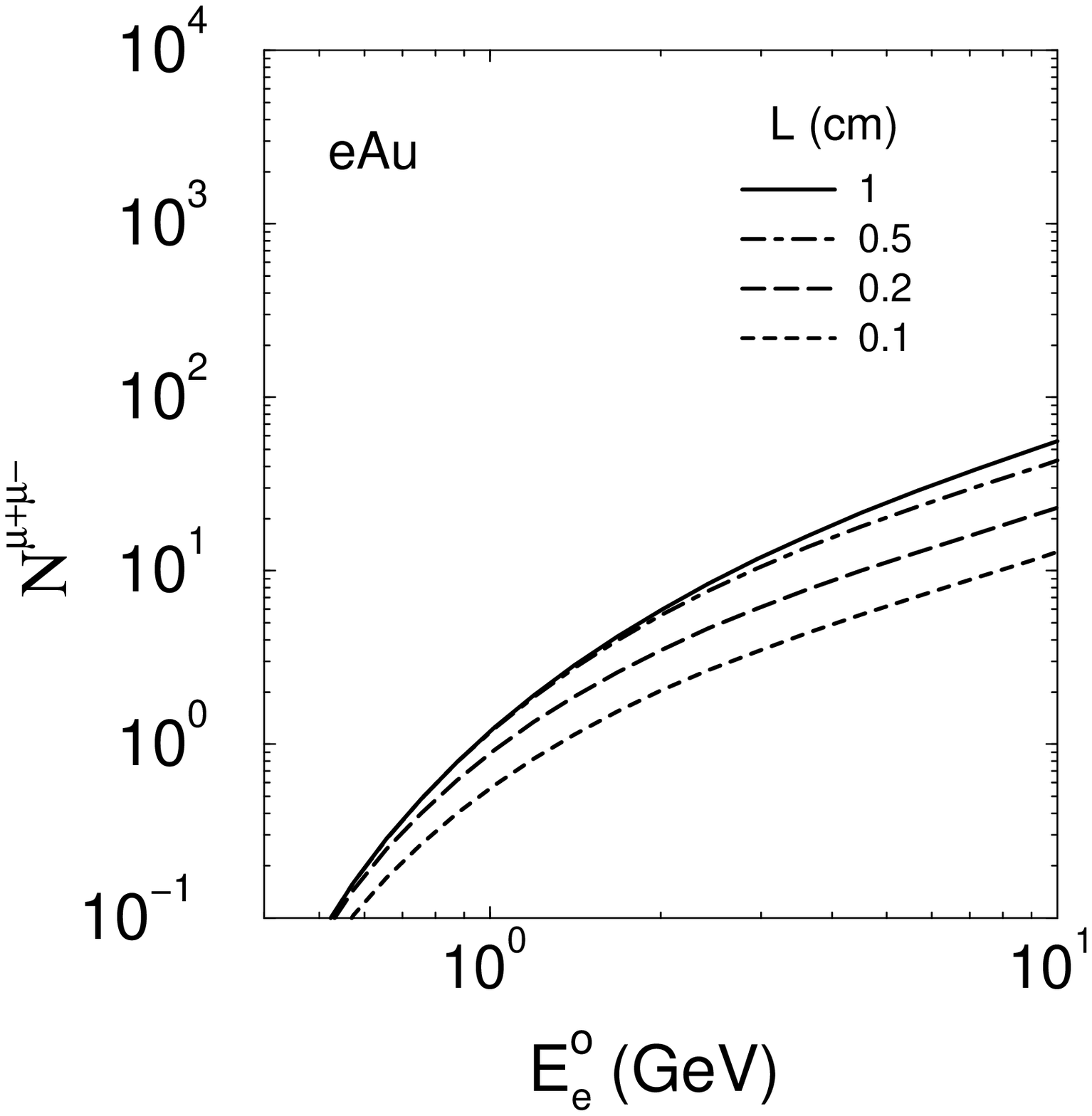}\qquad\qquad
      \includegraphics[width=0.3\columnwidth]{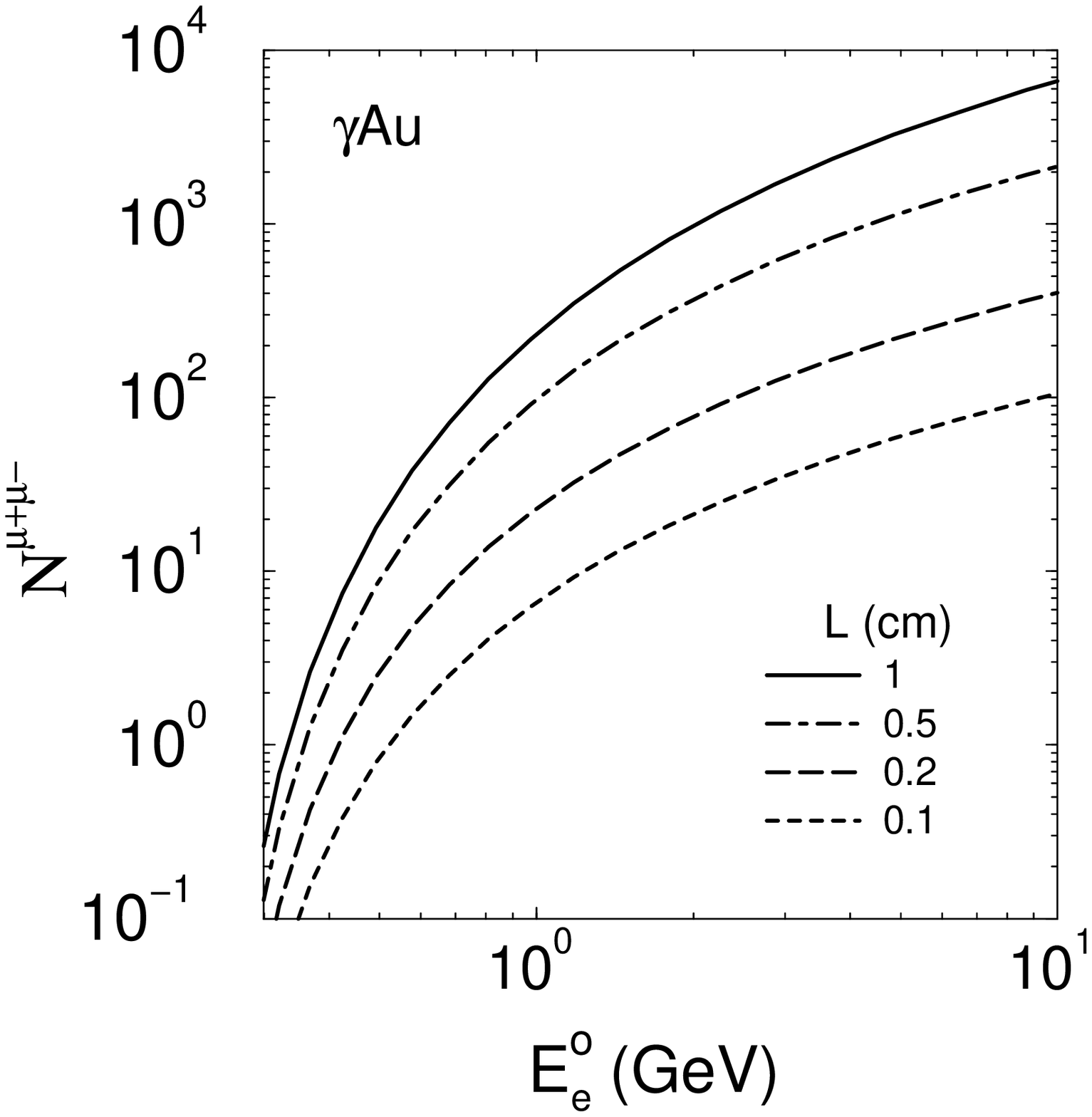}\qquad\qquad
   \caption{\small{
   The yields of dimuons in $e\,A$ (left panel) and $\gamma\,A$ (right panel)
   interactions
   as a function of primary electron energy $E_e^0$,
   calculated by Eqs.~(\protect\ref{E4})
   and (\protect\ref{E6}), respectively.
\label{Fig:7} }}
  \end{figure}

Results of calculations of the dimuon yields $N^{\mu^+\mu^-}$
predicted by Eqs.~(\ref{E4}) and (\ref{E6}) as a function of the
primary electron
energy $E_e^0$ are exhibited in Fig.~\ref{Fig:7}.
The total number of dimuons produced in $e\,A$ interactions is about 1 and
60 for $E_e^0=1$ and 10~GeV, respectively for thick target.
The yield increases
with the thickness of the target.
Some saturation sets in at
$L\ge L_{\rm sat.}\simeq 0.5$~cm.
This means that the dimuons are essentially produced in
a narrow region of the target mostly because of the large energy loss.
It seems to be natural, because the saturation length $L_{\rm sat.}$
is close to the scale parameter $l_0$ in Eq.~(\ref{E42}).

The number of dimuons in $\gamma\,A$ interactions is about 200 and
6000 for $E_e^0=1$ and 10~GeV, respectively, for target thickness of 1~cm.
The dependence of the total yield on the target thickness is rather strong
due to the large main free path of the bremsstrahlung photons.

\begin{figure}[h!]
   \includegraphics[width=0.3\columnwidth]{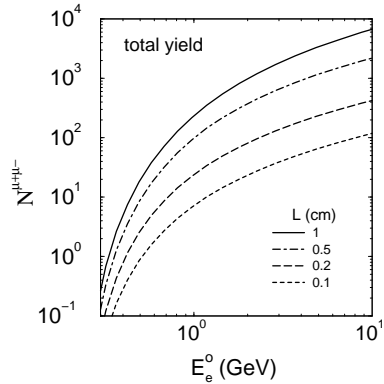}
   \caption{\small{
   The total yield of dimuons in  interactions
   of relativistic electrons with the gold target as a function
   of the primary electron
   energy $E_e^0$ and target thickness $L$.
\label{Fig:8} }}
  \end{figure}

The dimuon yield in $\gamma\, A$ interactions considerably exceeds
the corresponding yield in $e\,A$ interactions. This excess
strongly depends on the target thickness. Therefore,
in Fig.~\ref{Fig:8} we present the total dimuon yield in interactions
of relativistic electrons with a gold target which is a sum
of the two above contributions as a function of the primary electron
energy $E_e^0$ and target thickness $L$.
For thick targets it practically coincides with
the result shown in Fig.~\ref{Fig:7}~(right panel).

\section{Brief comment on hadronic processes}

Finally, let us consider the hadronic sources of muons in interactions of
 the laser-wakefield accelerated electrons with the thick target.
 The main source of muons here is the photo and/or
electro production and subsequent decay of
$\pi^\pm$ mesons  in the elementary processes
\begin{eqnarray}
 \gamma+p&\to& \pi^+ + n,\qquad
 \gamma+n\to \pi^-+p~,\label{E8a}\\
 \gamma+N&\to& \pi^+ + \pi^- +N~, \label{E8b}
\end{eqnarray}
where $\gamma$ stands for real bremsstrahlung photons (photoproduction) or
virtual photons (electroproduction). The production of $K$ mesons by 1 GeV electrons
is strongly suppressed because of energy arguments, however, at higher energies
they contribute to the muon production similarly to the pions.
Charged pions decay via
\begin{eqnarray}
\pi^+\to\mu^++\nu_\mu,\qquad \pi^-\to\mu^-+\bar \nu_\mu~
\label{E9}
\end{eqnarray}
with a probability of 99.99\%.
The cross sections of the reactions (\ref{E8a}) and (\ref{E8b})
for $E_\gamma\sim1$~GeV are around 100~$\mu$b
(see, for example,~\cite{SS}). Pion photoproduction in $\gamma\,A$ reactions
is proportional to $A^{2/3}$.
This means that the cross section
of the muon production at heavy target nuclei by
 $\gamma\,A\to \pi^\pm X\to \mu^\pm+\nu_\mu(\bar\nu_\mu) X$
 reactions exceeds the dimuon
photoproduction, considered in previous section, by a factor of $20-30$.
Taking into account the large pion absorption cross section
 at heavy target nuclei by strong interaction processes
 this means
 that the produced pions will be essentially absorbed in a thick target.
  The target must be thin enough to get a well collimated muon beam.
  This condition reduces the muon yield considerably.
  For example, for a dozen~$\mu$m thick gold target
 the expected yield of $\mu^\pm$ muons is less than 1 event for the
 above considered electron bunch.
 Therefore, the direct production of dimuons in electromagnetic interactions
 seems to be a more favorable mechanism.

 \section{summary}

 In summary we have considered muon pairs production by GeV electrons,
 created by a
 laser wakefield accelerator, impinging on a thick high-$Z$ target.
 We estimated the effectiveness of a such a source of muon pairs.
 For  a 1 cm thick gold target, $1.25\times10^8$ electrons
 in a 20 pC bunch with energy of 1 (10)~GeV
 in the initial state produce about $2\times10^{2}$ ($6\times10^{3}$)
 dimuons with pair energies centered at 1~GeV.
 To get $10^6$ dimuons from the muon factory one needs
  $10^{10}-10^{11}$ primary electrons in a bunch.
 Such intensities with power of 100~J seem to be quite realistic
 in near future, requiring ultra-high intensity laser pulses with
 efficient acceleration mechanisms.
 Cooling of the heat load in the target material analog
 to Ref.~\cite{Grosse} may be an option for higher repetition rates.
 The produced muons, unlike electrons and hadrons, penetrate
 the target material without suffering
 noticeable scattering and absorption.
 Thus the configuration of a laser driven electron accelerator
 and thick high-$Z$ target may serve as an all-optics table top device for
 muon pair production.
 The produced $\mu$ mesons may be used in studying various
 aspects of muon and neutrino physics and to be considered
 as an important  step towards investigations of more complicated
 electron induced elementary processes.

\acknowledgments
The authors appreciate S.V.~Bulanov, T.E. Cowan, and T.Zh.~Esirkepov
for fruitful discussions.


\begin{thebibliography}{30} 

\bibitem{C2}
W.P. Leemans {\it et al.}, Nature Physics {\bf 2}, 696 (2006).

\bibitem{Tajima1979}
 T.~Tajima and J.M.~Dawson,
 Phys. Rev. Lett. {\bf 43}, 267 (1979).

\bibitem{C3}
  S.~Karsch {\it et al.},
  New J.\ Phys.\  {\bf 9}, 415 (2007).

\bibitem{C4}
  J.~Osterhoff {\it et al.},
  Phys.\ Rev.\ Lett.\  {\bf 101}, 085002 (2008).

\bibitem{Vector}
  A.~I.~Titov and T.~S.~H.~Lee,
  Phys.\ Rev.\  C {\bf 66}, 015204 (2002).

\bibitem{Scalars}
A.~Donnachie and Yu.S.~Kalashnikova,
Phys. Rev. C {\bf78}, 064603 (2008).

\bibitem{Oh2001}
  Y.~S.~Oh, A.~I.~Titov, and T.~S.~H.~Lee,
  Phys.\ Rev.\  C {\bf 63}, 025201 (2001).

\bibitem{Strangeness}
B.~Saghai, J.-C. David, B. Julia-Diaz, and T.-S.H. Lee,
Eur. Phys. J. {\bf A 31}, 512 (2007).

\bibitem{Geer1998}
  S.~Geer,
  Phys.\ Rev.\  D {\bf 57}, 6989 (1998)
  [Erratum-ibid.\  D {\bf 59}, 039903 (1999)].

\bibitem{Geer2002}
  S.~Geer,
  J.\ Phys.\ G {\bf 29}, 1485 (2003).

\bibitem{ISS}
  A.~Bandyopadhyay {\it et al.}  [ISS Physics Working Group],
  arXiv:0710.4947 [hep-ph].

\bibitem{LFV}
  W.~J.~Marciano, T.~Mori, and J.~M.~Roney,
  Ann.\ Rev.\ Nucl.\ Part.\ Sci.\  {\bf 58}, 315 (2008).


\bibitem{Mu_g2}
  G.~W.~Bennett {\it et al.}  [Muon g-2 Collaboration],
  Phys.\ Rev.\  D {\bf 73}, 072003 (2006).

\bibitem{Bulanov}
  S.~V.~Bulanov, T.~Esirkepov, P.~Migliozzi,
  F.~Pegoraro, T.~Tajima, and F.~Terranova,
  Nucl.\ Instrum.\ Meth.\  A {\bf 540}, 25 (2005).

\bibitem{NT2002}
 K.~Nakashima and H.~Takabe,
 Phys. Plasmas {\bf 9}, 1505 (2002).

\bibitem{Karsch1999}
 S.~Karsch {\it et al.},
 Laser Part. Beams, {\bf 17}, 565 (1999).
\bibitem{Ruf2009}
  M.~Ruf, G.~R.~Mocken, C.~Muller, K.~Z.~Hatsagortsyan, and C.~H.~Keitel,
  Phys.\ Rev.\ Lett.\  {\bf 102}, 080402 (2009).

\bibitem{Muller2008}
  C.~Muller, C.~Deneke, and C.~H.~Keitel,
  Phys.\ Rev.\ Lett.\  {\bf 101}, 060402 (2008).

\bibitem{Raf}
  I.~Kuznetsova, D.~Habs, and J.~Rafelski,
  Phys.\ Rev.\  D {\bf 78}, 014027 (2008).

\bibitem{AB}
A.I.~Akhiezer and V.B.~Berestetsky,
Quantum Electrodynamics [Interscience,  1965].

\bibitem{Roche1971}
 G.~Roche, C.~Ducos, and J.~Prortol,
 Phys. Rev. A {\bf 5}, 2403 (1972).

\bibitem{Schiff}
  L.~I.~Schiff,  Phys.\ Rev.\  {\bf 83}, 252 (1951).

\bibitem{BA1972}
  E.~Borie and H.~Arenhoevel,
  Z.\ Phys.\  {\bf 255}, 459 (1972).

\bibitem{HEEA1}
T.~Tajima and  G.~Mourou,
Phys.\ Rev.\ ST Accel.\ Beams {\bf 12}, 051302 (2002).

\bibitem{HEEA2}
  G.~A.~Mourou, T.~Tajima, and S.~V.~Bulanov,
  Rev.\ Mod.\ Phys.\  {\bf 78}, 309 (2006).

\bibitem{HEEA3}
  N.~Kirby {\it et al.},
  Phys.\ Rev.\ ST Accel.\ Beams {\bf 12}, 051302 (2009).

\bibitem{Sch1997}
P.L.~Schkolnikov, A.E.~Kaplan, A.~Pukhov,
and J.~Meyer-ter-Vehn,
Appl. Phys. Lett. {\bf 71}, 3471 (1997).

\bibitem{Babichev}
A.P.~Babichev {\it et al}.,
Physical Values. Hand-book, (Eds). I.S.~Grigoriev and E.Z.~Meilikhov.
Energoiatomizdat, 1991 (in Russian).

\bibitem{BrData}
S.M.~Seltzer and M.J.~Berger,
At. Data Nucl. Data Tables,   {\bf 35}, 345 (1986).


\bibitem{SS}
  S.~Schadmand,
  Pramana {\bf 66}, 877 (2006)
  [arXiv:nucl-ex/0505023].

\bibitem{Grosse}
  J.~Klug {\it et al.},
  Nucl.\ Instrum.\ Meth.\  A {\bf 577}, 641 (2007).


\end{thebibliography}
\end{document}